\documentclass[prb,aps,twocolumn,showpacs]{revtex4}
\usepackage{dcolumn}
\usepackage{graphicx}
\usepackage{amssymb}
\begin{document}

\title{Metal adatoms on graphene and hexagonal boron nitride: \\
Towards the rational design of self-assembly templates}

\author{Oleg V. Yazyev}
\altaffiliation{Present address: Department of Physics, University of California, Berkeley, CA 94720, USA}
\email[E-mail: ]{yazyev@civet.berkeley.edu}
\affiliation{Institute of Theoretical Physics, Ecole Polytechnique 
F\'ed\'erale de Lausanne (EPFL), CH-1015 Lausanne, Switzerland}
\affiliation{Institut Romand de Recherche Num\'erique en Physique
des Mat\'eriaux (IRRMA), CH-1015 Lausanne, Switzerland}
\author{Alfredo Pasquarello}
\affiliation{Institute of Theoretical Physics, Ecole Polytechnique 
F\'ed\'erale de Lausanne (EPFL), CH-1015 Lausanne, Switzerland}
\affiliation{Institut Romand de Recherche Num\'erique en Physique
des Mat\'eriaux (IRRMA), CH-1015 Lausanne, Switzerland}

\date{\today}

\begin{abstract}
Periodically corrugated epitaxial graphene and hexagonal boron nitride ($h$-BN)
on metallic substrates are considered as perspective templates for the 
self-assembly of nanoparticles arrays. By using first-principles calculations, 
we determine binding energies and diffusion activation barriers of metal 
adatoms on graphene and $h$-BN. The observed chemical trends can be understood in 
terms of the interplay between charge transfer and covalent bonding involving the 
adatom $d$ electrons. We further investigate the electronic effects of the metallic 
substrate and find that periodically corrugated templates based on graphene in combination
with strong interactions at the metal/graphene interface are the most suitable
for the self-assembly of highly regular nanoparticle arrays.
\end{abstract}

                
\maketitle

\section{INTRODUCTION}

Two-dimensional graphene and hexagonal boron nitride ($h$-BN) are attracting
considerable attention due to their extraordinary physical properties and
perspective technological applications.\cite{Geim07,Katsnelson07} Epitaxial 
single layers of graphene and $h$-BN can be grown via the chemical vapor deposition 
route on a large variety of metallic substrates.\cite{Oshima97} In many cases, 
the epitaxial layers of these two materials are extremely well-ordered but reveal 
long-wavelength periodic corrugations, or Moir\'e patterns, resulting from the 
lattice constant mismatch.\cite{Laskowski07,Coraux08,deParga08,Martoccia08} Such 
superlattices are considered as promising templates for the chemical self-assembly of 
periodic arrays of nanoparticles, with perspective applications in ultra-high density 
information storage, catalysis, sensing, etc.\cite{Berner07} The potential of this 
approach has already been affirmed by the successful production of regular arrays of 
nanoclusters with narrow size distributions.\cite{NDiaye06,Zhang08,NDiaye09,Donner09} 

One attractive property of this approach to nanoscale self-assembly is the high 
degree of customization. Indeed, the space of allowed chemical compositions involves 
three degrees of freedom: {\it (i)} deposited metallic nanoparticles formed by virtually
any metal from the periodic table;  {\it (ii)} a monolayer of either graphene or $h$-BN;
{\it (iii)}  metallic surfaces, one of the many (111) fcc or (0001) hcp surfaces of forth 
or fifth row late transition metals. Understanding the roles these three factors play
in the self-assembly process is of paramount importance for the rational design of 
nanoparticle arrays with novel properties and functions. 

In this work, we aim at understanding the chemical trends in the binding and diffusion 
of individual metal adatoms, the initial step which largely predetermines the
overall self-assembly process. In particular, by using first-principles calculations 
we perform a systematic study of the adatom binding and diffusion as well as of the 
electronic properties of the metal adatoms upon deposition. First, we focus on the 
chemical trends observed for metal adatoms varying horizontally and vertically in the 
periodic table, when absorbed on free-standing graphene and $h$-BN. Then, we model the 
electronic effects of periodically corrugated epitaxial layers deposited on two metallic 
substrates representing the limiting cases of strong and weak interactions at the 
interface. 

The present paper is organized as follows. In Sec.\ \ref{methods} we describe 
our first-principles methodology and the adopted models. Section~\ref{adatoms} 
discusses the binding and the diffusion of adatoms on free-standing graphene 
and $h$-BN.  Section~\ref{substrate} is devoted to the investigation 
of substrate effects. The conclusions are drawn in  Sec.\ \ref{conclusions}.

\section{METHODS}\label{methods}

In our approach, the electronic structure is described through the use of 
the Perdew-Burke-Ernzerhof (PBE) exchange-correlation density functional
within density functional theory.\cite{Perdew96}
The ultrasoft pseudopotentials \cite{Vanderbilt90} used in the present study treat the 
$d$-electrons of all transition metals as valence electrons. The semi-core $sp$-states 
were also treated explicitly in the case of early third row elements (K--Fe). 
The one-electron valence wave functions and the electron density were 
described by plane-wave basis sets with kinetic energy cutoffs of 30~Ry and 300~Ry, 
respectively.\cite{Pasquarello92} All calculations were performed using the 
spin-unrestricted formalism.  We used the \textsc{pwscf} plane-wave pseudopotential 
code of the \textsc{quantum-espresso} distribution.\cite{QE} The convergence of the 
results with respect to the simulation parameters was systematically verified. 

The constructed models are based on periodic two-dimensional 3$\times$3 supercells 
in combination with 4$\times$4 ${\mathbf k}$-point meshes. Potential energy surfaces 
for the free-standing graphene and $h$-BN systems were studied by relaxing all 
atomic positions with the initial coordinates of the adatom set to the high-symmetry 
positions shown in Fig.~\ref{fig1}. This procedure allowed us to determine both the 
potential energy surface minima and the transition state configurations connecting
the neighboring equivalent minima, since symmetry was retained during the relaxation. 
The validity of this approach was verified through nudged elastic band calculations.\cite{Mills95}
For specific cases, we could compare our results with those of other 
investigations,\cite{Chan08,Sevincli08} finding good agreement. 

We investigated the effects of local electronic-structure variations due 
to the presence of metallic substrates by considering the case of Co 
adatoms on epitaxial graphene and $h$-BN supported by lattice-matched Ni(111) 
and Cu(111) substrates. These metallic substrates correspond to regimes of strong 
and weak monolayer-substrate interactions, respectively.\cite{Giovannetti08} 
Under realistic conditions these metals do not produce Moir\'e pattern due to 
the small lattice mismatch.\cite{Oshima97,Nagashima94,Gamo97,Li09} However, through 
models involving such structures, we could study electronic effects induced 
by the Moir\'e superlattice using sufficiently small simulations cells and 
without introducing large lateral strains. A very similar methodology was 
also used in Ref.~\onlinecite{Laskowski07}.  The considered model calculations 
consist of 3$\times$3 two-dimensional slabs composed of 4 atomic planes of 
metal with both surfaces covered by graphene or $h$-BN in order 
to avoid spurious dipole-dipole interactions. The slab configurations 
were fully optimized, including the distance between the metal layers.\cite{Kadas06} 


\section{Adatoms on suspended graphene and $h$-BN}\label{adatoms}

\begin{figure}
\includegraphics[width=8.2cm]{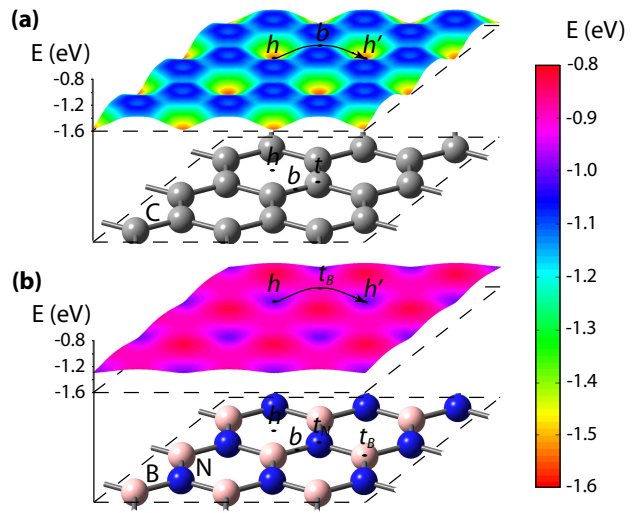}
\caption{\label{fig1} (Color online) 
Potential energy surfaces of the Co adatom on freestanding (a) graphene and (b) $h$-BN
obtained by constraining the in-plane position of the Co adatom on a 6$\times$6 points mesh in the unit cell.
The same energy scale is used in both plots and the energy is referred with respect 
to that of an isolated Co atom. The high-symmetry positions are indicated with labels.
The arrows show the lowest energy diffusion pathways connecting the local minima, 
$h$ sites, through the transition state configurations, the $b$ sites and $t_{\rm B}$ sites 
in the case of graphene and $h$-BN, respectively.
}
\end{figure}

First, let us compare the potential energy surfaces (PES) for a representative case of a Co 
adatom on free-standing graphene and $h$-BN (Fig.~\ref{fig1}). For both monolayers 
the PES minima correspond to the hollow ($h$) sites. 
Interestingly, in the case of Co as well as for most of the other 
metal adatoms the distortions of the graphene and $h$-BN lattices are very weak. 
Although the two PES involving the Co adatom are qualitatively 
similar, the binding energy on graphene are substantially higher (1.60~eV vs.\ 1.03~eV).
The stronger binding to graphene is a systematic feature which can be 
understood considering that the chemisorption of metal adatoms is governed by 
the following attractive contributions: (i) strong covalent bonding, and (ii) moderate 
electrostatic interaction as a result of charge transfer between the adatom and the 
monolayer. While graphene is a semimetal, $h$-BN is an insulator with a band gap of 
$\approx$6~eV.\cite{Arnaud06} The availability of low-energy electronic 
states in graphene results in a more efficient interaction with the adatom states 
and allows for charge transfer between the graphene layer and the adatom, thus 
introducing an electrostatic component.
The preferential binding to the $h$ sites is very common, although we find a number of 
exceptions, especially among the heavy transition metals.
In particular, for Pd, Ir and Pt, the bridge ($b$)
site is the lowest energy position on graphene. The nitrogen on-top sites 
($t_{\rm N}$) correspond to the PES minima on $h$-BN for V, Ni, Pd, Ir
and Pt. 

\begin{figure}\includegraphics[width=8.2cm]{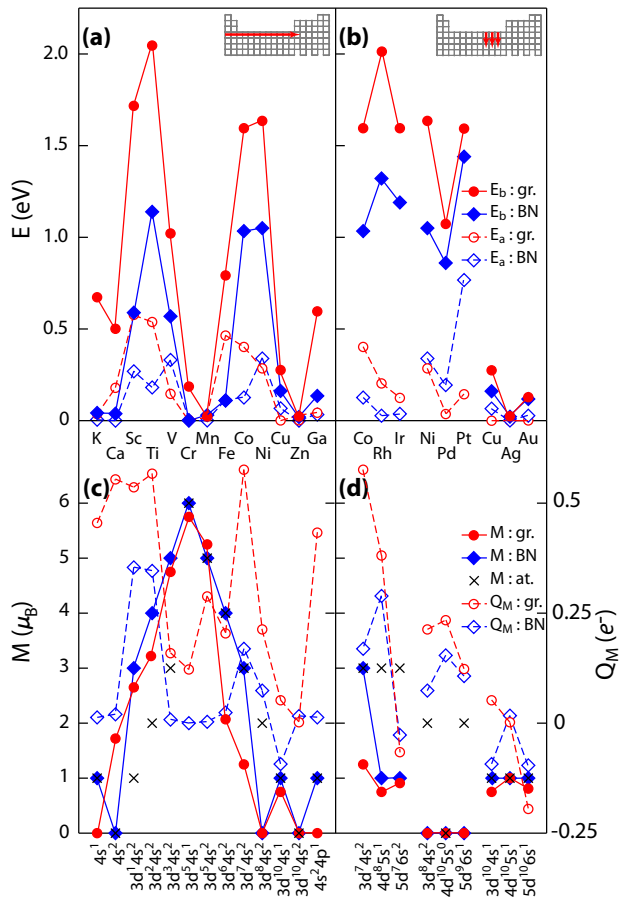}
\caption{\label{fig2} (Color online) 
(a) Horizontal and (b) vertical chemical trends in the binding energies $E_{\rm b}$ and 
the diffusion activation barriers $E_{\rm a}$ of metal adatoms on graphene and $h$-BN. 
Insets show the  concerned parts of the periodic table: K--Ga in (a) and Co--Ir,
Ni--Pt, Cu--Au in (b). (c) Horizontal and (d) vertical chemical trends in the magnetic
moments $M$ and the atomic charges $Q_{\rm M}$ of the adsorbed metal adatoms in their 
lowest energy configurations. Crosses refer to the magnetic moments of isolated atoms.
Electronic configurations of the isolated atoms are given at the bottom of the figure
for reference.
}
\end{figure}

The lowest energy pathways connecting the neighboring local minima involve transition 
states at $b$ and $t_{\rm B}$ sites for the Co adatom on graphene and $h$-BN, respectively. 
The diffusion activation barriers are again higher for graphene, 0.40~eV vs.\ 0.13~eV for $h$-BN. 
However, this trend is not systematic across the periodic table and very often the 
diffusion on $h$-BN is characterized by higher activation barriers [cf.\ Fig.~\ref{fig2}(a,b)]. 
Notably, the calculated activation barrier in the case of $h$-BN agrees well with the 
experimental value of 0.14$\pm$0.03~eV for the diffusion of Co adatoms on 
$h$-BN deposited on Ni(111).\cite{Auwarter02}

Figure~\ref{fig2}(a) shows the horizontal trend in the binding energies and the 
diffusion activation barriers across the third row of the periodic table. 
Both graphene and $h$-BN show double-peaked features with maximum binding energies
at Ti and Ni which have $d^2$ and $d^8$ electronic configurations, respectively, 
while the very weak binding of Cr and Mn corresponds to the situation of half-filled 
$d$-shells. Strong binding energies with magnitudes up to 2.08~eV (1.14~eV) for 
graphene ($h$-BN) indicate the contribution from covalent binding involving $d$
electrons. However, the fact that the atomic magnetic moments due to the partially 
filled $d$-shells are largely preserved upon binding 
[Fig.~\ref{fig2}(c)] shows that graphene and $h$-BN act as weak ligand fields with 
respect to the metal adatoms. Large magnetic moments associated to half-filled $d$-shell 
elements correspond to large values of spin splitting of $d$-electron states and, thus, 
to their reduced participation to the covalent binding. 
The other elements in the first (second) half of the transition metal series
tend to give enhanced (reduced) magnetic moments upon binding. To elucidate 
the origin of this behavior, we performed the L\"owdin population analysis 
\cite{Lowdin50} and found that the adsorption of metal adatoms leads to the partial promotion 
of electrons from the $s$- to the $d$-shell thus leading to the observed changes of 
the magnetic moments. 

Such trends depending on the partial $d$-shell filling appear to be generic since
a very similar behavior was found for metal-benzene molecular complexes\cite{Pandey01}
and even for strongly bound substitutional impurities in graphene.\cite{Krasheninnikov09,Boukhvalov09,Santos10}
The binding energies on $h$-BN as well as the activation barriers on both monolayers
follow exactly the same qualitative trend within the transition metal series (Sc--Zn).
Outside of this region (K, Ca and Ga) only graphene is able to bind adatoms strongly 
since the interactions are then governed by the sole electrostatic contribution resulting
from charge transfer. This contrast is well illustrated by comparing the L\"owdin atomic 
charges $Q_{\rm M}$ of metal adatoms on the two monolayers [Fig.~\ref{fig2}(c)].  

Vertical trends in the periodic table are illustrated in Fig.~\ref{fig1}(b) for the 
late transition and coinage metal adatoms. In general, the binding energies tend to increase when moving 
down the columns of the periodic table. The reductions can be associated with the changes of
the lowest energy adsorption sites discussed above. The largest binding energy of 2.02~eV
is found for Rh on graphene while the highest activation barrier of 0.77~eV corresponds 
to Pt on $h$-BN. At variance, the diffusion of Pd on graphene and of Rh and Ir
on $h$-BN are characterized by very low diffusion barriers ($<$0.05~eV) combined 
with high binding energies ($>$1~eV). Inert coinage metals have very low binding 
energies and diffusion barrier, with minimum values for Ag. Similar trends 
were also observed for carbon on coinage metal surfaces.\cite{Yazyev08} 
Even on graphene, Ag shows practically no charge transfer as this element is located at 
the point of crossover between electron donating (Cu) and electron accepting (Au) behavior 
[Fig.~\ref{fig2}(d)]. The hole-doping of graphene which is otherwise difficult to achieve 
by chemical means was recently demonstrated in Au deposition experiments.\cite{Gierz08}
Interestingly, we find that also strongly binding Ir adatoms act as electron acceptors.

It is worth stressing that the presently used theory does not provide a correct 
description of weak van der Waals interactions. This might quantitatively 
affect the results, especially when the considered adatom does not give rise to 
strong covalent bonding and/or electrostatic interactions.

\section{Effect of metallic substrate}\label{substrate}

\begin{figure*}
\includegraphics[width=11cm]{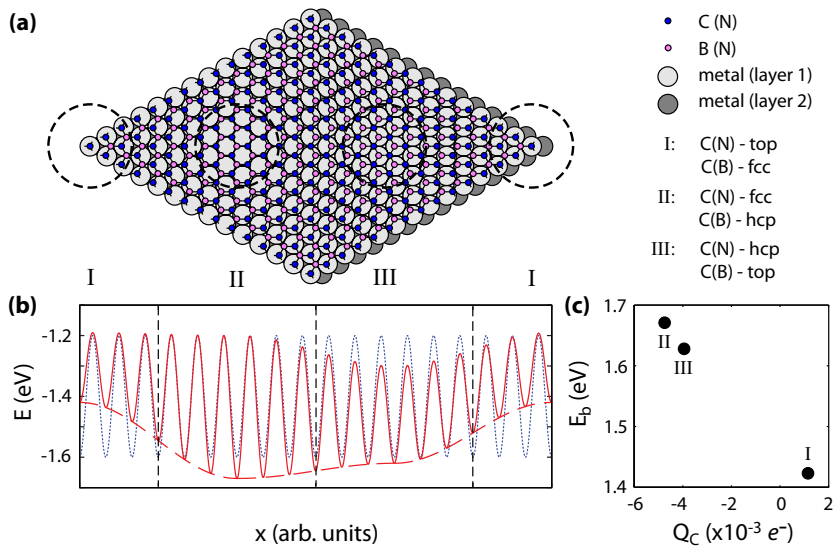}
\caption{\label{fig3} (Color online) 
(a) Schematic illustration of the Moir\'e unit cell with three binding regions
indicated. (b) Schematic representation of the potential energy 
surface (solid line) for the diffusion of the adatom moving along the horizontal direction
of the unit cell shown in (a), based on calculations for Moir\'e domains I--III. 
The actual local binding energies (dashed line) 
and diffusion barriers (oscillation amplitudes) correspond to the case of the Co adatom 
on graphene supported by Ni(111). The dotted line refers to the case of the Co adatom on
free-standing graphene. (c) Binding energy $E_{\rm b}$ of the Co adatom versus the
charge transfer per carbon atom $Q_{\rm C}$ from the metal to graphene for the three 
Moir\'e domains.
}
\end{figure*}

On the surfaces of forth and fifth row transition metals, epitaxial graphene 
and $h$-BN produce Moir\'e patterns due to the mismatch between the lattice constants of the 
substrate and of the monolayer [cf.\ Fig.~\ref{fig3}(a)]. The local shifts of the
monolayer lattice with respect to the substrate lattice across the Moir\'e unit cell 
result in long-wavelength modulations of the PES which drive the 
self-assembly of periodic arrays of nanoparticles. These modulations are due to
variations in the monolayer-substrate distance and to changes in the local electronic 
structure.\cite{deParga08} 
Below, we focus on the role of local electronic structure variations in the diffusion of individual 
metal adatoms on metal-supported graphene and h-BN. Collective effects due to the local 
rehybridization induced by few-atom clusters have been considered elsewhere.\cite{Feibelman08,Feibelman09}
In addition, the distance modulations alone can lead to only minor
curvature-induced effect on the adatom binding energies since the estimated height
variations are actually rather small ($\Delta h$$<$1.5~\AA~\cite{Laskowski07,Martoccia08}) 
compared to the size of the Moir\'e unit cell ($a$$\sim$3~nm).\cite{footnote}

In order to understand the effect of the local electronic structure variations, we study 
the binding of Co adatoms on epitaxial graphene and $h$-BN supported by lattice-matched 
Ni(111) and Cu(111) substrates. These two cases model the regimes of strong 
and weak monolayer-substrate interactions, respectively.\cite{Giovannetti08} We stress that under 
realistic conditions these metals do not produce Moir\'e pattern due to the small lattice 
mismatch.\cite{Oshima97,Nagashima94,Gamo97,Li09} The three principal 
Moir\'e domains [I--III in Fig.~\ref{fig3}(a)] are modeled by introducing an artificial 
lateral shift between the monolayer and the substrate. Further details of this procedure
are given in Sec.\ \ref{methods}.

The calculated binding energies and activation barriers for Co adatoms on graphene$|$Ni(111) 
show appreciable variation across the three Moir\'e domains [cf.~Table~\ref{tab1}]. 
The resulting PES [schematically depicted in Fig.~\ref{fig3}(b)]
can be viewed as a modulation of the PES of free-standing
graphene [Fig.~\ref{fig1}(a)] by the long-wavelength perturbations resulting from the periodicity 
of the Moir\'e superlattice. The Moir\'e domain I, which is characterized by the shortest
graphene-substrate distance (2.23~\AA),\cite{Bertoni05,Karpan07,Yazyev09} shows the lowest 
binding energy while the opposite is true for region II showing the largest graphene-substrate 
distance (4.11~\AA). 
However, Moir\'e domain III presents a short graphene-substrate distance (2.49~\AA)
in combination with a high binding energy [Table~\ref{tab1}], indicating that there is 
no clear correlation between these quantities. At variance, a clear correlation can be
established between the binding energies of the Co adatoms and the local charge transfer 
per carbon atom ($Q_{\rm C}$) from the metal to graphene
[cf.\ Fig.~\ref{fig3}(c)]. This result can be intuitively 
understood: the charge transfer from the metal to graphene reduces the charge transfer from the 
Co adatom to the template surface and, thus, the electrostatic component of the
binding energy.
Very recently a similar electrostatic effect has also been found for adatoms on ultrathin 
oxide films.\cite{Giordano08} For Co adatoms on $h$-BN$|$Ni(111), we found the
same tendency, although the local variations of the binding energy are smaller in this case. 
On a weakly binding substrate such as the Cu(111) surface, the PES remains practically 
unchanged across the Moir\'e unit cell. 

\begin{table}[b]
\caption{\label{tab1} Comparison of binding energies $E_{\rm b}$ (in eV) and diffusion 
activation barriers $E_{\rm a}$ (in eV) of the Co adatom on free-standing graphene (gr.) 
and $h$-BN with corresponding epitaxial monolayers deposited on Ni(111) and Cu(111). 
The different positions of the monolayer atoms refer to three distinct local regions of 
the Moir\'e pattern shown in Fig.~\ref{fig3}. 
}
\begin{ruledtabular}
\begin{tabular}{lcccccc}
                &      & $E_{\rm b}$  &   &   &  $E_{\rm a}$ &   \\ \cline{2-4} \cline{5-7}
 Moir\'e domain     &  I   & II  & III  &  I   & II  & III \\ 
\hline
 gr.              &       & 1.60  &       &       & 0.40  &        \\  
 gr.$|$Ni(111)    & 1.42  & 1.67  & 1.62  & 0.23  & 0.47  & 0.32   \\    
 gr.$|$Cu(111)    & 1.66  & 1.67  & 1.67  & 0.45  & 0.45  & 0.45   \\  
 $h$-BN           &       & 1.03  &       &       & 0.13  &        \\
 $h$-BN$|$Ni(111) & 0.86  & 0.89  & 0.88  & 0.10  & 0.14  & 0.14   \\
 $h$-BN$|$Cu(111) & 1.17  & 1.17  & 1.17  & 0.16  & 0.20  & 0.20  
\end{tabular}
\end{ruledtabular}
\end{table}

The Moir\'e domains associated with higher binding energies are expected to show larger thermal
populations of adatoms, and will thus act as nucleation centers for nanoparticles. 
Therefore, larger variations of the binding energies across the Moir\'e template will favor 
more ordered nanoparticle arrays with narrower size distributions. We conclude that
the templates made of graphene in combination with strongly binding metal surfaces 
(e.g. Ir, Rh, Ru) are more promising for the self-assembly of metal nanoparticle arrays. 

\section{CONCLUSIONS}\label{conclusions} 

We systematically studied the potential energy surfaces of metal adatoms on graphene and 
$h$-BN across the periodic table, and clarified the role of the metallic 
substrate in adatom diffusion. The present results are well understood within a simple picture 
involving covalent and electrostatic interactions. Our work formulates general principles 
required for the rational design of self-assembly templates based on epitaxial graphene and 
$h$-BN which can find applications in ultra-high density information storage, catalysis and 
sensing, and so on.

\section*{ACKNOWLEDGMENTS}

We acknowledge fruitful discussions with H.~Brune, P.~Buluschek, R.~Decker, A.~Lehnert, and S.~Rusponi.  
The calculations were performed at the CSCS.


\begin{thebibliography}{33}
\bibitem{footnote}
{We assume spherically curved regions of diameter $d$$=$$a/2$. The estimated 
curvature radius $R$$\approx$$a^2/(32\Delta h)$$\geq$2~nm is too large to 
produce any sizable effect on the local binding energies.\cite{Ribeiro05}}
\end{thebibliography}

\end{document}